\documentclass{aa}

\usepackage{graphicx}
\usepackage[varg]{txfonts}
\usepackage{hyperref}

\begin{document}

\title{
    Finding the dispersing siblings of young open clusters
}
\subtitle{Dynamical traceback simulations using \textit{Gaia}~DR3}

\author{
    E.~Vaher\inst{\ref{inst:Lund}}
    \and D.~Hobbs\inst{\ref{inst:Lund}}
    \and P.~McMillan\inst{\ref{inst:Lund},\ref{inst:Leicester}}
    \and T.~Prusti\inst{\ref{inst:ESTEC}}
}
\institute{
    Lund Observatory, Division of Astrophysics, Department of Physics, Lund University, Box 43, 22100 Lund, Sweden\label{inst:Lund}
    \and School of Physics \& Astronomy, University of Leicester, University Road, Leicester, LE1~7RH, UK\label{inst:Leicester}
    \and European Space Agency (ESA), European Space Research and Technology Centre (ESTEC), Keplerlaan 1, 2201 AZ Noordwijk, The Netherlands\label{inst:ESTEC}
}

\abstract{
	Stars tend to form in clusters, but many escape their birth clusters very
    early. Identifying the escaped members of clusters can inform us about the
    dissolution of star clusters, but also about the stellar dynamics in the
    galaxy. Methods capable of finding escaped stars from many clusters are
    required to fully exploit the large amounts of data in the \textit{Gaia}
    era.
}
{
    We present a new method of identifying escaped members of nearby clusters
    and apply it to ten young clusters.
}
{
    We assumed the escaped stars were close to the cluster in the past and
    performed traceback computations based on the \textit{Gaia}~DR3 radial
    velocity subsample.
    For each individual star, our method produces a probability estimate that
    it is an escaped member of a cluster, and for each cluster it also
    estimates the field star contamination rate of the identified fugitives.
}
{
	Our method is capable of finding fugitives that have escaped from their
    cluster in the last few ten million years.
    In many cases the fugitives form an elongated structure that covers a
    large volume.
}
{
	The results presented here show that traceback computations using
    \textit{Gaia}~DR3 data can identify stars that have recently escaped
    their cluster.
    Our method will be even more useful when applied to future \textit{Gaia}
    data releases that contain more radial velocity measurements.
}


\maketitle

\section{Introduction}

Stars do not form in isolation: most are born in dense star-forming regions
\citep{2003ARA&A..41...57L}.
Not all stars that form together are gravitationally bound.
Furthermore, the loss of mass when the leftover gas is ejected releases
many more stars into the Galactic field very early in the cluster's life.
Even the clusters that are still bound after that will eventually fall apart
due to the gravitational ejection of stars, close encounters with giant
molecular clouds, or interactions with the Milky Way tidal field
\citep[see][for a review]{2019ARA&A..57..227K}.
Most clusters dissipate within $10-100\,\mathrm{Myr}$, leaving many stars
dispersed in the Galactic field
\citep[e.g.,][]{2003ARA&A..41...57L, 2006A&A...445..545P}.

Identifying the escaped members of open clusters could enable further research
on many topics, of which we only list a handful here.
Finding exoplanet systems around young stars allows models of planet formation
and migration to be refined \citep[see, e.g.,][]{2007MNRAS.375...29A}, and
identifying stars that have escaped young clusters can enable more such systems
to be found.
\citet{2021arXiv210602050K} performed simulations of a Milky
Way-like galaxy and found that it might be possible to use stellar streams from
disintegrating open clusters in the Galactic disk to constrain the parameters
of the Galactic bar.
\citet{2022ApJ...925..214D} have recently proposed a method of determining
cluster ages based on the orientation of their tidal tails.
Although their method in its current state is, strictly speaking, only
applicable to clusters on circular orbits, an improved version for realistic
noncircular orbits might be suitable for providing age estimates for many
clusters, independent of stellar evolution models.
\citet{2022MNRAS.517.3613K} have even suggested that the asymmetry
between the number of stars in the leading and trailing tidal tails of clusters
could be used as a test to compare Newtonian and Milgromian gravitation.

Stars from the same formation region can cover a large volume in position space,
even if they have very similar velocities, so distinguishing escaped cluster
stars from the much more numerous unrelated field stars can be like searching
for a needle in a haystack.
True young stellar groups are expected to be chemically homogeneous, so strong
chemical tagging could contribute to the solution.
However, determining detailed chemical abundances requires very high quality
spectra, and the number of stars for which such spectra are available is rather
low.
The European~Space~Agency's \textit{Gaia} mission has published in its third
data release \citep[\textit{Gaia}~DR3;][]{2021A&A...649A...1G} the full
six-dimensional phase space information for $\sim$34~million sources all over
the sky.
The availability of this data means that a method capable of finding cluster
escapees based on kinematics rather than spectroscopy is very appealing.
Even if a kinematic method is not capable of perfectly separating escaped
cluster stars from field stars, it can still be valuable if it can produce
a candidate list with a small enough field contamination rate, which could then
be used as a starting point for further analysis 
\citep[see, e.g.,][]{2021AJ....162..197B}.

The convergent point method
\citep[see, e.g.,][for recent applications]{2019A&A...621L...2R, 2019A&A...627A...4R, 2020A&A...638A...9R}
assumes that stars originating from the same cluster have similar velocities.
The observed proper motion of a star can be compared with the proper motion it
would have if its velocity components were replaced with the velocity components
of the cluster.
If the difference is too large, then the star is not considered to be associated
with the cluster, but if the difference is small enough, then it might be.
The convergent point method is computationally inexpensive, so it can be
applied to a large number of clusters.
Furthermore, stars for which radial velocity measurements are not feasible can
still be included in the analysis.
However, the simplest implementation of the method makes it difficult to
recognize stars with velocities too similar to the cluster velocity, for which
there is no reason to suspect that they were any closer to the cluster in the
past than they are now.

\citet{2019MNRAS.489.3625C} introduced a method that can provide
probabilities that stars belong to associations as well as the ages of these
associations.
They assume associations are composed of one or more components that can
be described as six-dimensional Gaussians in phase space with some specific age.
The time evolution of these components in a gravitational potential can be
computed, and the likelihood that observed stars belong to a component can then
be used to fit its parameters.
Although \citet{2019MNRAS.489.3625C} successfully applied their method to the
$\beta$~Pictoris moving group, the fact that they assigned each component a
specific expansion age means their method is not appropriate for studying open
cluster tidal tails, which consist of stars that are escaping the cluster
continuously.

\citet{2021A&A...647A.137J} performed $N$-body computations in a Galactic
potential to find out how the tidal tails of the open cluster Hyades should look
in the phase space and then searched for stars with compatible astrometry.
Their method proved to be successful at identifying stars in the Hyades tidal
tails up to $800\,\mathrm{pc}$ away from the cluster itself.
However, the $N$-body computations required by their method make it rather
resource-intensive, so for studying a large number of clusters a simpler method
is appealing.

Recently \citet{2021A&A...645A..84M} searched for extended comoving stellar
populations around ten nearby young open clusters.
Their method involves making a hard cut in projected proper motion space
followed by applying a clustering algorithm in position space.
They were able to find vast stellar halos that are comoving with the clusters.
Therefore, the parameter values of their clustering algorithm seem to be well
justified empirically, but it is not simple to evaluate or interpret them on
physical grounds.
Furthermore, the use of a clustering algorithm means that they can only find
the stars where the population density is relatively high.

The idea of finding escaped cluster members by looking for stars that have been
close to the cluster in the past is well motivated on physical grounds.
The requirement that the current position of an escapee relative to a cluster
must be consistent with both its current relative velocity and escape time
introduces correlations in phase space that allow us to reject many unrelated
field stars that might be broadly consistent with the escapee population in
every phase space coordinate individually.
This idea has been utilized by \citet{2020ApJ...900...14F} and
\citet{2020MNRAS.495.3104S}, who both looked for stars that have escaped from
the Orion Nebula Cluster.
However, the Orion Nebula Cluster is young enough that both
\citet{2020ApJ...900...14F} and \citet{2020MNRAS.495.3104S} ignored the
Milky Way gravity entirely and assumed the stars move in straight lines with
unchanging speed, so their methods are not applicable to older clusters.

In this paper we introduce a method that uses traceback computations to identify
stars that have been close to the cluster in phase space in the past.
Section~\ref{sect:method} provides the details of our method and describes how
it can be used to estimate individual membership probabilities for the escaped
stars and also to estimate the field star contamination rate of the identified
population. 
In Sect.~\ref{sect:results} we outline the results of applying our method to
the same ten clusters that \citet{2021A&A...645A..84M} studied.
We show that our method is indeed able to identify escaped cluster stars that
currently cover a large volume in phase space, and we provide a brief comparison
with \citet{2021A&A...645A..84M}.
In Sect.~\ref{sect:discussion} we vary some of the parameters of our analysis
and show that our results are not sensitive to the exact parameter value
choices we have made.
Finally, Sect.~\ref{sect:summary} summarizes our conclusions and discusses the
strengths and weaknesses of our method.

\section{Method}\label{sect:method}

\subsection{General idea and definitions}\label{subsect:general_idea}

We looked for stars that have escaped from a cluster by performing traceback
computations and identifying the stars that have been in a small phase space
volume around the cluster in the not too distant past.
The search volumes we considered are spherical in both position and velocity
space, with radii of $r_p= 15\,\mathrm{pc}$ and 
$r_v= 3\,\mathrm{km\,s^{-1}}$, respectively.
In the following we call a volume with such radii around a point in phase
space the ``neighborhood'' of that point.
The position space radius of our neighborhoods is quite similar to the typical
tidal neighborhood of an open cluster \citep[$\sim 10\,\mathrm{pc}$; see, e.g.,
Table~2 of][]{2021A&A...645A..84M}, and the velocity space radius is likewise
similar to the typical velocity dispersion of stars originating from the same
open cluster
\citep[a couple of $\mathrm{km\,s^{-1}}$; see again Table~2 of][]{2021A&A...645A..84M}.
Values much smaller would be too restrictive and exclude even many stars still
gravitationally bound to the cluster, whereas values much larger should be
expected to include too many stars that have never been gravitationally bound.
We provide further empirical justification for the radii we have chosen in
Sect.~\ref{subsect:neighborhood_size}.

We performed 100 Monte Carlo tracebacks by sampling the current phase space
coordinates of the stars based on the reported \textit{Gaia} uncertainties and
correlations.
We interpreted the fraction of tracebacks in which a star is in the cluster
neighborhood at any time up to some maximum traceback time as the star's
``fugitive probability,'' $p_f$, and the fraction of traceback simulations a
star starts out in a cluster's neighborhood as the star's ``membership
probability,'' $p_m$.
We point out that $p_f \geq p_m$ by definition.
We also defined a ``probability threshold,'' $p_\mathrm{min}$, and used it to
divide the stars into three disjoint subsets:

\begin{itemize}
    \item Stars with $p_f \geq p_m > p_\mathrm{min}$, which we call cluster
        ``members.''
        These are presently inside the cluster neighborhood or very close to it,
        which can be seen even without any traceback computations.
    \item Stars with $p_f > p_\mathrm{min} \geq p_m$, which we call
        ``fugitives.''
        These are not current members, but the traceback computations reveal
        them to have been in the cluster neighborhood in the past with a large
        enough probability to be of interest.
    \item Stars with $p_\mathrm{min} \geq p_f \geq p_m$, which we consider to
        be unrelated to the cluster.
\end{itemize}

We found it useful enough for our purposes to adopt the value
$p_\mathrm{min}=0.1$, but a higher or a lower value could be chosen to increase
the purity or the completeness of the sample, respectively.
It is convenient that the probability threshold could be adjusted without having
to perform a new set of traceback computations.

Our definitions do mean that in principle some stars we label as fugitives might
qualify as members if a different value of $p_\mathrm{min}$ were used.
We show empirically in Sect.~\ref{subsect:p_min_discussion} that this is not a
concern in practice.

\subsection{Initial selection of stars}\label{subsect:initial_selection}

Traceback computations require full knowledge about the kinematics of the
included stars.
We are therefore naturally limited to the \textit{Gaia}~DR3 radial velocity data
set with five-parameter astrometry.
We corrected for the \textit{Gaia}~DR3 parallax zero-point offset using the
procedure outlined by \citet{2021A&A...649A...4L} and discarded sources with
corrected parallaxes smaller than $1\,\mathrm{mas}$.
We then converted the coordinates of the stars to galactocentric cylindrical
coordinates using the parameter values listed in
Table~\ref{tab:solar_parameters} and retained the sources that satisfy the
conditions listed in Table~\ref{tab:galactocentric_selection}.
The number of such sources is 4\,238\,086.

We remark that the standard deviation of the difference between the inverted
parallax of the final sample with and without the parallax offset correction is
$8.3\,\mathrm{pc}$ and the standard deviation of the radial velocity
uncertainty is $3.4\,\mathrm{km\,s^{-1}}$.
We can make a rough estimate that for traceback times beyond the ratio of
the aforementioned values $2.4\,\mathrm{Myr}$ the typical uncertainty in the
distance of a star caused by the uncertainty of the radial velocity is larger
than the distance difference that would be caused by neglecting the parallax
offset entirely.
Because this timescale is much shorter than the age of even the youngest cluster
in our selection, we conclude that our results are not sensitive to the details
of parallax offset correction.

In our Monte Carlo tracebacks, any stars with large uncertainties in astrometry
or radial velocity will have a large spread in their computed past phase space
coordinates, which ensures that stars with too uncertain coordinates have low
fugitive probabilities.
Our choice of $p_\mathrm{min}=0.1$ is therefore implicitly applying a
quality cut to the data, and we can afford to not apply any explicit conditions
to the uncertainties of the astrometric parameters.
Furthermore, in Sect.~\ref{subsect:rv_uncertainty_dominance} we show that the
dominant source of uncertainty for us is the radial velocity, so commonly used
quality criteria based on \textit{Gaia} astrometry, such as the renormalized
unit weight error (RUWE) value, are not useful for us.

\begin{table*}[t]
	\begin{center}
		\caption{
            Adopted parameters of the solar orbit.
        }
		\begin{tabular}{l c c c}
			\hline\hline
			Parameter & Symbol & Value & Reference \\
			\hline
			Sun's height from midplane & $z_\sun$ & $14\,\mathrm{pc}$ & 1\\
			Sun's peculiar velocity & $(U_\sun,V_\sun,W_\sun)$ & $(11.10,12.24,7.25)\,\mathrm{km\,s^{-1}}$ & 2\\
			Sun's galactocentric distance & $R_\odot$ & $8.0\,\mathrm{km\,s^{-1}}$ & 3 \\
			Local circular velocity & $v_0$ & $220\,\mathrm{km\,s^{-1}}$ & 3\\
			\hline
		\end{tabular}
        \tablefoot{
            The parameters are consistent with the \texttt{MWPotential2014}
            gravitational potential we used for the traceback computations.
        }
		\tablebib{
            (1)~\citet{1997MNRAS.288..365B};
            (2)~\citet{2010MNRAS.403.1829S};			(3)~\citet{2015ApJS..216...29B}.
        }
		\label{tab:solar_parameters}
	\end{center}
\end{table*}

\begin{table*}
\centering
\caption{
\label{tab:galactocentric_selection}
Limits for the coordinates of the stars included in the initial traceback computation.}
\begin{tabular}{lccc}
\hline \hline
Parameter & Symbol & Minimum & Maximum \\
\hline
Galactocentric distance & $R$ & $7500.0\,$$\mathrm{pc}$ & $8500.0\,$$\mathrm{pc}$ \\
Galactocentric azimuth & $\phi$ & $170.0\,$$\mathrm{{}^{\circ}}$ & $190.0\,$$\mathrm{{}^{\circ}}$ \\
Height from galactic midplane & $z$ & $-350.0\,$$\mathrm{pc}$ & $350.0\,$$\mathrm{pc}$ \\
Galactocentric radial velocity & $v_R$ & $-50.0\,$$\mathrm{km\,s^{-1}}$ & $30.0\,$$\mathrm{km\,s^{-1}}$ \\
Galactocentric azimuthal angular velocity & $v_\phi$ & $-1.9\,$$\mathrm{{}^{\circ}\,Myr^{-1}}$ & $-1.3\,$$\mathrm{{}^{\circ}\,Myr^{-1}}$ \\
Vertical velocity & $v_z$ & $-20.0\,$$\mathrm{km\,s^{-1}}$ & $20.0\,$$\mathrm{km\,s^{-1}}$ \\
Heliocentric distance & $d$ & $0.0\,$$\mathrm{pc}$ & $1000.0\,$$\mathrm{pc}$ \\
\hline
\end{tabular}
\end{table*}

\subsection{Cluster coordinates}\label{subsect:clusterparameters}

In addition to stars our traceback simulations also include points representing
the clusters themselves.
We refined the initial guesses of the heliocentric Cartesian phase space
coordinates of a cluster by iteratively updating them to be the medians of the
coordinates of all the stars within the neighborhood of the previous guess until
the cluster coordinates converged.
Using median rather than the mean makes the cluster coordinates less sensitive
to possible field star contamination.
The nominal cluster coordinates were obtained by querying the 5 astrometric
parameters and radial velocity from SIMBAD \citep{2000A&AS..143....9W} with
\texttt{astroquery} \citep{2019AJ....157...98G} for the initial guess (see
Table~\ref{tab:rough_cluster_coordinates}) and taking the stellar coordinates
from \textit{Gaia}~DR3, only correcting for the global parallax zero-point
offset \citep{2021A&A...649A...4L}.
We Monte Carlo-sampled the uncertainty of the cluster coordinates using the same
iterative procedure with each set of Monte Carlo-sampled stellar coordinates
and with the nominal cluster coordinates being the initial guess.

\subsection{Field star contamination}

\begin{figure}
    \centering
    \includegraphics{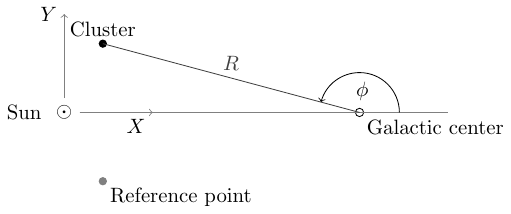}
    \caption{
        Galactocentric cylindrical coordinates and the heliocentric Cartesian
        coordinates.
        The $z$ (galactocentric) and $Z$ (heliocentric) coordinates are almost
        perpendicular to the plane of the figure.
        The Sun is located at
        $(R,\phi,z)=(8000\,\mathrm{pc},180\degr,14\,\mathrm{pc}),$
        and the Galactic center is located at
        $(X,Y,Z)=(8000\,\mathrm{pc},0\,\mathrm{pc},0\,\mathrm{pc})$. 
    }
    \label{fig:coordinates}
\end{figure}

Some of the stars that have been in a cluster's neighborhood might be unrelated
field stars.
We estimated the number of such interlopers by counting the number of stars in
the neighborhood of a reference point.
This comparison is only meaningful if the reference point is carefully chosen.
Assuming that the galactic field is axisymmetric and its spatial density is
symmetric with respect to the galactic midplane, we can choose a reference point
that has the same field star density in its neighborhood as the cluster at all
times.
Let the present-day cylindrical galactocentric phase space coordinates of the
cluster be $(R, \phi, z, v_R, v_\phi, v_z)$ (see also
Fig.~\ref{fig:coordinates}) and let $(R', \phi',..., v'_z)$ be the present-day
coordinates of the reference point.
The reference point should have the same galactocentric distance as the cluster
at all times so that the field star density in their neighborhoods would not
differ because of a galactic radial density gradient, so $R'=R$ and $v_R'=v_R$.
To avoid differences caused by the galactic vertical density gradient at all
times we should choose either $z'=z$ and $v_z'=v_z$ or $z'=-z$ and $v_z'=-v_z$.
Our assumption of an axisymmetric field requires us to choose $v_\phi'=v_\phi$,
but allows for all values of $\phi'$.
However, \textit{Gaia} being magnitude-limited introduces a heliocentric
distance bias to the observed field star density, which restricts us to three
possible reference points, with galactocentric cylindrical position space
coordinates $(R, \phi, -z)$, $(R, 360\degr-\phi, z)$ and
$(R, 360\degr-\phi, -z)$.
We used as the reference point the last of the three because it is the farthest
from the cluster, which minimizes the chance that there are cluster fugitives in
its neighborhood.
In each Monte Carlo traceback the initial coordinates of the reference point
were computed from the cluster coordinates sampled for that traceback.

The Sun is not exactly in the galactic midplane, so the heliocentric distance
of the chosen reference point and the cluster is not exactly the same.
However, the vertical offset of the Sun is small, and we ignored the distance
difference this offset causes.

We required members and fugitives to have $p_f > p_\mathrm{min}$.
Likewise, we ignored those stars for which the fraction of Monte Carlo
tracebacks where they entered the reference neighborhood did not exceed
$p_\mathrm{min}$.

\begin{table}
\centering
\caption{\label{tab:cluster_parameters}
Cluster ages and metallicities from \citet{2019A&A...623A.108B}.}
\begin{tabular}{lccc}
\hline \hline
Name & Cluster age & [Fe/H] & [Fe/H] reference \\
 & $\mathrm{Myr}$ &  &  \\
\hline
\object{Platais 9} & $78.3_{-4.0}^{+10.2}$ & 0.0 &  \\
\object{Messier 39} & $309.7_{-56.8}^{+74.8}$ & 0.0 & 1 \\
\object{$\alpha$ Per cluster} & $86.7_{-1.0}^{+1.2}$ & 0.14 & 1 \\
\object{NGC 2451 A} & $44.4_{-1.8}^{+1.7}$ & 0.0 & 2, 3 \\
\object{IC 2602} & $35.2_{-1.1}^{+1.3}$ & 0.0 & 1 \\
\object{NGC 2547} & $27.0_{-1.1}^{+1.1}$ & 0.0 & 2, 3 \\
\object{Blanco 1} & $94.4_{-6.7}^{+5.1}$ & 0.0 & 1 \\
\object{IC 2391} & $36.4_{-1.7}^{+1.9}$ & 0.0 & 1 \\
\object{NGC 2516} & $251.2_{-3.4}^{+2.9}$ & 0.0 & 1 \\
\object{Pleiades} & $86.5_{-2.4}^{+6.0}$ & 0.0 & 1 \\
\hline
\end{tabular}
\tablefoot{The cluster ages are median values of the Bayesian posteriors with 16th and 84th percentiles.
The metallicity values listed are the same that were used for fitting the cluster isochrones.
In the case of \object{Platais~9} the solar metallicity was assumed.}

\tablebib{(1) \citet{2016A&A...585A.150N}; (2) \citet{2017A&A...601A..70S}; (3) \citet{2017A&A...603A...2M}.}
\end{table}

\subsection{Traceback computations}

We performed traceback computations using a leapfrog integrator implemented in
\texttt{Gala} \citep{2017JOSS....2..388P}.
The gravitational potential we used was \texttt{BovyMWPotential2014}, which is
the implementation of \texttt{MWPotential2014} from \texttt{galpy}
\citep{2015ApJS..216...29B} in \texttt{Gala}.

We recorded the state of the computation with a timestep of
$0.25\,\mathrm{Myr}$, which is reasonably small compared to the timescale
dictated by our choice of neighborhood radii of
$15\,\mathrm{pc}\,/\,3\,\mathrm{km\,s^{-1}} \approx 5\,\mathrm{Myr}$.
The maximum traceback time that should be used is more difficult to judge.
On one hand, a traceback too short prevents us from finding fugitives that left
their cluster early, which suggests an individual traceback time for each
cluster based on its age.
On the other hand, longer traceback times mean larger uncertainties of the
computed phase space coordinates, which suggests a global maximum traceback time
beyond which the computations are uninformative.
Because it is simple to ignore the traceback results that are too far in the
past, we chose a single value of $100\,\mathrm{Myr}$ for all the clusters,
knowing that it is greater than most of the cluster ages in our sample (see
Table~\ref{tab:cluster_parameters}).
The fugitive probabilities we report, however, only consider the traceback
snapshots before the cluster age estimates by \citet{2019A&A...623A.108B},
listed in Table~\ref{tab:cluster_parameters}.
For the two clusters that are older than $100\,\mathrm{Myr},$ we used that as
the limit instead.
We show in Sect.~\ref{subsect:maximum_traceback_time} empirically
that longer traceback times are not justified with \textit{Gaia}~DR3 data.

Although a star can be strongly influenced by cluster gravity if it is close to
the cluster in phase space, our criteria for identifying members and fugitives
is only based on whether the star is in the cluster neighborhood or not.
The exact trajectory of the star within the neighborhood has no importance for
us, so attempting to include cluster gravity in our computations would be a
complication with no benefit.

For convenience, we excluded from the Monte Carlo tracebacks the stars that are
always far outside the neighborhood of every cluster and reference point.
We identified them by performing an initial traceback computation with a
traceback time of $100\,\mathrm{Myr}$ and with the initial selection of stars
described in Sect.~\ref{subsect:initial_selection}.
We retained the stars that are closer to any nominal cluster or reference point
at any timestep than a generous velocity space separation of
$10\,\mathrm{km\,s^{-1}}$ and time-dependent position space separation
of $50\,\mathrm{pc} + (1\,\mathrm{km\,s^{-1}})\cdot t$, where $t$ is
the traceback time.
The time-dependent radius was used here because uncertainty of the current
cluster velocity contributes to uncertainty of the cluster position in the past,
but an even larger constant position space radius would contain many more
unrelated stars in our selection.
We did not use a time-dependent radius in the Monte Carlo tracebacks because the
Monte Carlo procedure samples the cluster velocity uncertainty directly.

\section{Results}\label{sect:results}

\begin{table*}
\centering
\caption{\label{tab:probabilities}Membership and fugitive probabilities of the \textit{Gaia}~DR3 sources.}
\begin{tabular}{cccccc}
\hline \hline
Source ID & $\alpha$ & $\delta$ & Membership probability & Fugitive probability & Cluster \\
 & deg & deg &  &  &  \\
\hline
1344158844768366464 & 263.7829 & 40.52364 & 0.0 & 0.02 & Platais 9 \\
5226381982652636928 & 164.8539 & -73.0302 & 0.0 & 0.13 & Platais 9 \\
6006623127671951360 & 231.8738 & -37.7566 & 0.0 & 0.05 & Platais 9 \\
5809549996411193472 & 249.5263 & -67.1619 & 0.0 & 0.02 & Platais 9 \\
6227406539994517120 & 226.5135 & -25.3748 & 0.0 & 0.01 & Platais 9 \\
\hline
\end{tabular}
\tablefoot{Only the first five rows are displayed. The full table is available in machine-readable form at the CDS.}
\end{table*}

\begin{table*}
\centering
\caption{
    \label{tab:cluster_results}
    False positive rates, which are the medians of the false positive rates of
    the Monte Carlo tracebacks, shown with the 16th and 84th percentiles.
}
\begin{tabular}{lccc}
\hline \hline
Name & No. of fugitives & No. of members & False positive rate \\
 &  &  & $\mathrm{\%}$ \\
\hline
\object{Platais 9} & 99 & 32 & $5_{-5}^{+5}$ \\
\object{Messier 39} & 24 & 81 & $14_{-4}^{+14}$ \\
\object{$\alpha$ Per cluster} & 143 & 186 & $8_{-3}^{+4}$ \\
\object{NGC 2451 A} & 40 & 78 & $0_{-0}^{+11}$ \\
\object{IC 2602} & 47 & 125 & $7_{-2}^{+5}$ \\
\object{NGC 2547} & 40 & 102 & $0_{-0}^{+8}$ \\
\object{Blanco 1} & 19 & 122 & $0_{-0}^{+26}$ \\
\object{IC 2391} & 55 & 90 & $5_{-5}^{+6}$ \\
\object{NGC 2516} & 115 & 559 & $7_{-4}^{+5}$ \\
\object{Pleiades} & 60 & 395 & $11_{-4}^{+6}$ \\
\hline
\end{tabular}
\end{table*}

Table~\ref{tab:probabilities} is an excerpt that shows the \textit{Gaia}~DR3
source IDs, on-sky coordinates and membership and fugitive probabilities of a
few stars.
The full table that contains all stars with positive fugitive probabilities is
available at the CDS.
The numbers of cluster members and fugitives together with the false positive
rates at the cluster ages are summarized in Table~\ref{tab:cluster_results}.
We provide a more detailed description of the clusters below.

\begin{figure*}
    \centering
    \includegraphics[width=17cm]{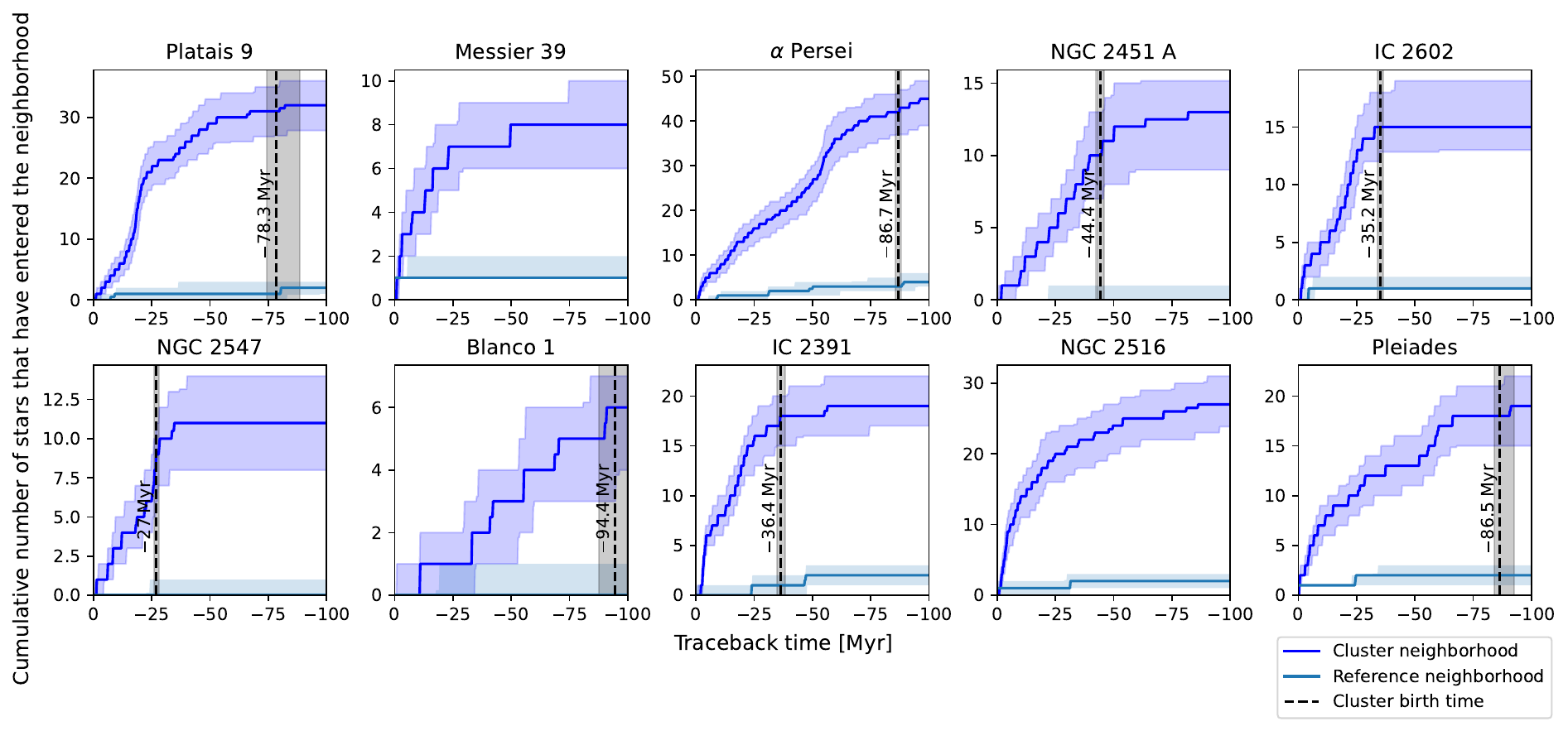}
    \caption{
        Comparison of the cumulative number of stars in the reference
        neighborhood (light blue) and the cumulative number of stars that
        enter the cluster neighborhood (dark blue).
        The solid lines show the median counts over the 100 Monte Carlo
        traceback computations.
        The colored areas show the 16th and 84th percentiles of the cumulative
        counts.
        The vertical dashed lines and gray bands mark the cluster ages and
        uncertainties according to \citet{2019A&A...623A.108B}, except for the
        two clusters that are much older than the maximum traceback time of
        $100\,\mathrm{Myr}$.
        Many more stars enter the cluster neighborhood than the reference
        neighborhood, suggesting that there is indeed a population of escaped
        cluster stars that the traceback computations are able to find.
    }
    \label{fig:cumulative_counts}
\end{figure*}

\subsection{Detection of the fugitive populations}

For each Monte Carlo traceback, we tracked the cumulative number of stars that
have entered the neighborhood of a cluster or a reference point as a function of
time.
Figure~\ref{fig:cumulative_counts} shows the medians of these cumulative counts
together with the 16th and 84th percentiles after having removed all stars with
$p_f \leq p_\mathrm{min}$.
It can be seen that the cumulative counts in the cluster neighborhoods are
significantly larger than in the reference neighborhoods.
This clear and consistent pattern means that each cluster is indeed accompanied
by a population of escaped stars, and furthermore our traceback method is able
to identify many of the fugitives.

\subsection{Cluster ages and maximum traceback time}
\label{subsect:maximum_traceback_time}

Because there should be no fugitives older than the cluster itself, we expect to
see many stars entering the cluster neighborhood in our traceback computations
up to the moment of its birth, and none beyond that.
The traceback times at which the cumulative count plots in
Fig.~\ref{fig:cumulative_counts} flatten out could then be used as estimates
of the clusters' age.
We also show the cluster age estimates by \citet{2019A&A...623A.108B} in
Fig.~\ref{fig:cumulative_counts} and it can be seen that in many cases there
indeed seems to be good agreement between their cluster age estimate and the
flattening of the cumulative count curve, (see, e.g., \object{NGC 2451 A},
\object{IC 2602}, or \object{IC 2391}).
However, a similar flattening can be observed for \object{Messier 39} and
\object{NGC 2516}, which should be much older than our maximum traceback time
of $100\,\mathrm{Myr}$.
This can be explained by the fact that the further back in time our traceback
computations go, the larger the uncertainties of the computed phase space
coordinates become.
This, in turn, means the further in the past a star escaped its cluster, the
less likely we are to see it enter the cluster neighborhood in our computations.
\citet{2021arXiv210602050K}, who performed simulations of a Milky Way-like
galaxy, find that even with perfect knowledge of the large-scale gravitational
potential, stochastic encounters with giant molecular clouds still prevent
accurate tracebacks of stars beyond timescales of a couple hundred million
years.
We should expect our useful traceback time to be even shorter.
The cumulative count plots of \object{Messier 39} and \object{NGC 2516} indeed
suggest that the timescale on which the computed phase space coordinates of the
stars become too uncertain to be useful for us is tens of millions of years,
which happens to be similar to the age of most of the clusters in our sample.
Disentangling the resulting change in the cumulative counts' slopes from that
caused by the finite age of the clusters is not feasible, so our computations
cannot be used to estimate the cluster ages, they can only be used to provide
lower bounds.
There are no clusters in our sample for which the number of stars having been
within its neighborhood grows significantly beyond traceback times further than
the cluster age estimates from \citet{2019A&A...623A.108B}, so our lower
bounds for the cluster ages are consistent with their isochrone ages.

From the discussion above it is clear that extending the traceback computations
beyond $100\,\mathrm{Myr}$ would not be useful.
However, smaller current-day phase space coordinate uncertainties would result
in smaller past coordinate uncertainties, so longer traceback times could be
used with future \textit{Gaia} data releases.
Furthermore, it might become possible to provide traceback age estimates for
some of the clusters in our sample without having to modify our traceback
method itself, or at the very least the lower bounds for the cluster ages could
be made more restrictive.

\begin{figure*}
	\centering
    \includegraphics[width=17cm]{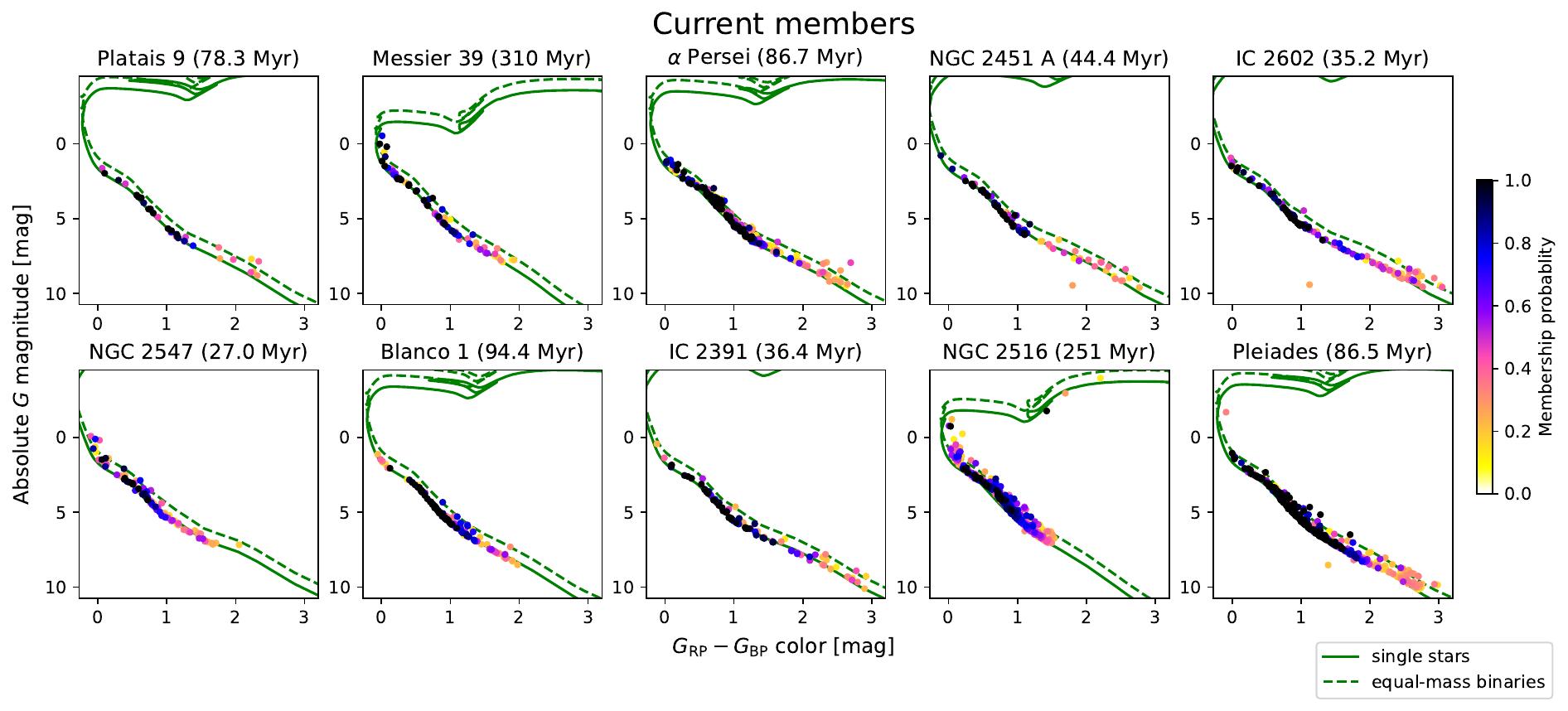}
    \includegraphics[width=17cm]{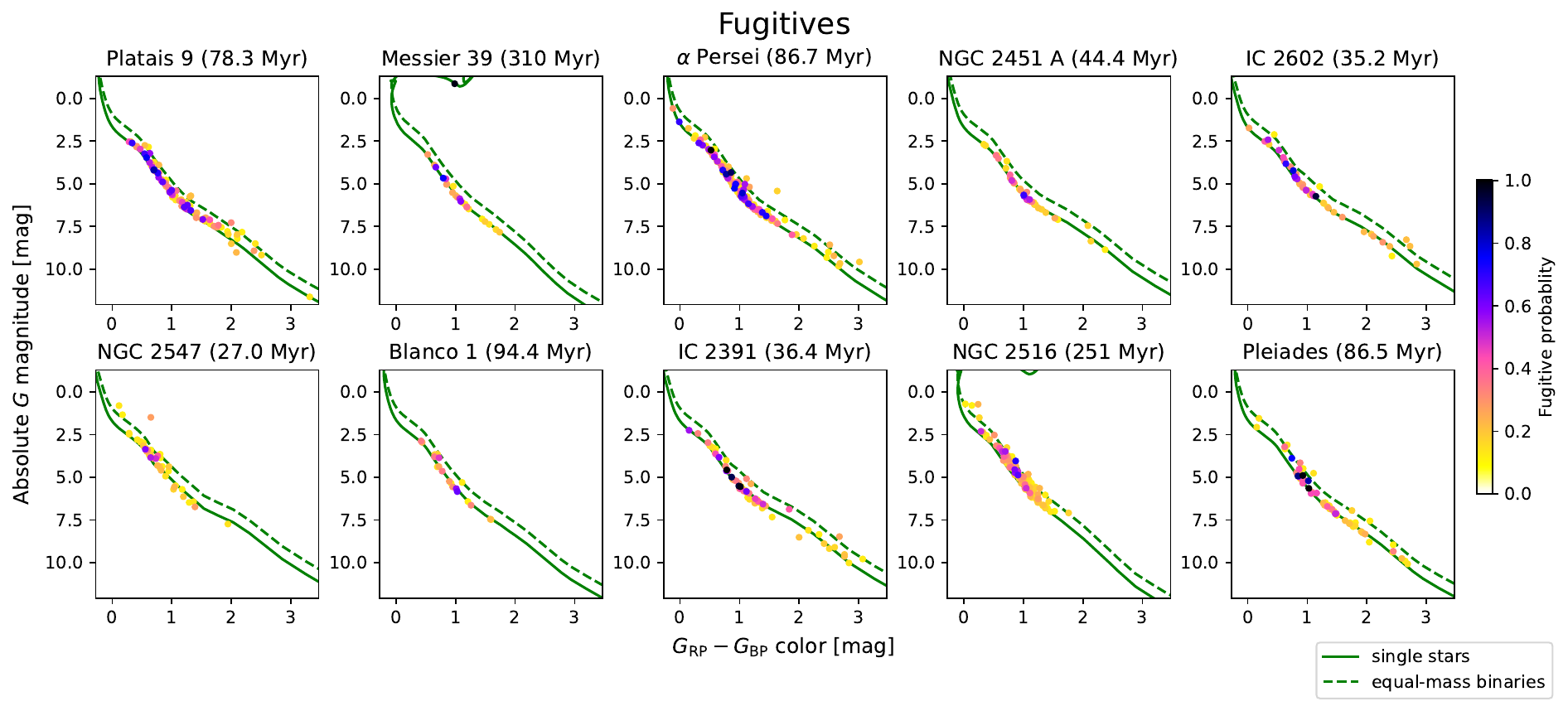}
    \caption{
        HR diagrams of the cluster members (top ten) and fugitives (bottom ten)
        together with PARSEC isochrones.
        The isochrone ages in the panel titles are taken from
        \citet{2019A&A...623A.108B}.
        The metallicities of the isochrones are listed in
        Table~\ref{tab:cluster_parameters}.
        The fugitives agree with the isochrones just as well as the members,
        other than a handful of outliers with very low fugitive probabilities.
        The presence of such outliers is to be expected, given the estimated
        false positive rates listed in Table~\ref{tab:cluster_results}.
    }
	\label{fig:HRdiagram}
\end{figure*}

\subsection{Hertzsprung–Russell diagrams}

Figure~\ref{fig:HRdiagram} compares the members and fugitives we found with
PARSEC isochrones
\citep{2012MNRAS.427..127B, 2014MNRAS.444.2525C, 2015MNRAS.452.1068C} on the
Hertzsprung–Russell (HR) diagram.
The parameters of the isochrones are listed in
Table~\ref{tab:cluster_parameters}.
The cluster members outnumber field stars in the current cluster neighborhood
owing to their much larger phase space density.
It is therefore not surprising that the observed HR diagrams of the members
match the isochrones very well.
There are a handful of clear outliers, but they all have low membership
probabilities.
It is more remarkable that the observed HR diagrams of the fugitives also
match the same isochrones because the fugitives are spread over a large volume
of phase space that contains many unrelated field stars.
There are again a handful of outliers with low fugitive probabilities, but the
presence of a small number of outliers is to be expected based on the estimated
false positive rates listed in Table~\ref{tab:cluster_results}.

\begin{figure*}
	\centering
	\includegraphics[width=17cm]{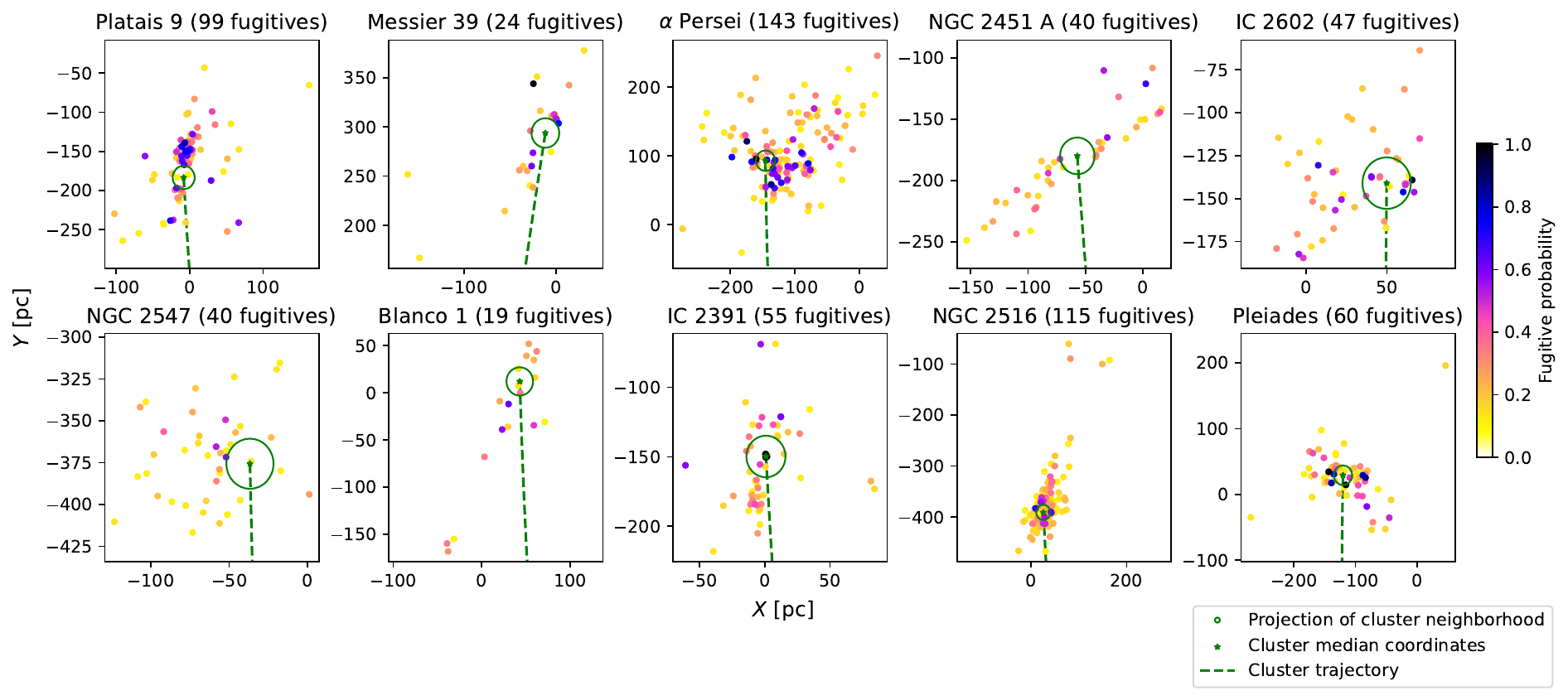}
	\caption{
	    Current-day distribution of the fugitives projected on the galactic
	    plane.
	    Some of the high-probability fugitives are currently far from their
	    clusters.
	}
	\label{fig:XYcoordinates}
\end{figure*}

\subsection{Spatial distribution of fugitives}

We now briefly compare the spatial distribution of the cluster fugitives
with the structures found by \citet{2021A&A...645A..84M}.
Figure~\ref{fig:XYcoordinates} can be compared with their Fig.~A.2.
In most cases, the fugitives we find seem to form a stream with the inner (i.e.,
closer to the Galactic center) arm leading and the outer arm trailing, but this
is not true for all clusters.
In the cases where we do see a convincing stream, \citet{2021A&A...645A..84M}
also see a stream with the same orientation.
The fact that the structures we find are so similar validates both our and
\citet{2021A&A...645A..84M} methods and also confirms that the identified
structures are real.

\subsubsection{Platais 9}

\citet{2021A&A...645A..84M} found a prominent stream associated with this
cluster, and the fugitives we find also tend to form a stream with the same
orientation and a similar extent.
The part of the stream between us and the cluster seems to be more prominent
than the part behind the stream, but that is most likely due to observational
biases.
We also find a couple of stars with relatively high fugitive probabilities
quite far from the stream.
The method of \citet{2021A&A...645A..84M} includes applying a clustering
algorithm in position space, which prevents them from finding such stars.

\subsubsection{Messier 39}

The fugitives seem to form a stream with the same orientation as the stream
found by \citet{2021A&A...645A..84M} and a similar extent.
However, the number of high-probability fugitives is rather low.

\subsubsection{$\alpha$ Persei cluster}

This is the cluster with the largest number of fugitives, but the fugitives do
not seem to form a stream.
It is not obvious why that would be the case. 
According to \citet{2019A&A...623A.108B}, the \object{$\alpha$ Per cluster} has
a very similar age as \object{Pleiades}, but despite that, the distribution of
the fugitives is very different.
The structure \citet{2021A&A...645A..84M} report is elongated, but
relatively thick compared to the streams they found around many of the other
clusters.

\subsubsection{NGC 2451 A}

The fugitives form a clear stream that has quite a large angle with respect to
the trajectory of the cluster.
The angle and the rough extent of the stream we see are very similar to the
stream found by \citet{2021A&A...645A..84M}.

\subsubsection{IC 2602}

Once again the fugitives form a stream that is similar to the stream found by 
\citet{2021A&A...645A..84M}, but differently from the other clusters with a
clear stream in this case the cluster is not centered on the stream but rather
somewhat displaced laterally.
This feature is also visible in the results of \citet{2021A&A...645A..84M}.

\subsubsection{NGC 2547}

\citet{2021A&A...645A..84M} do not see a stream around this cluster, instead
they see an extended corona.
The distribution of fugitives seems to be consistent with the findings of
\citet{2021A&A...645A..84M}, but the number of fugitives is low.

\subsubsection{Blanco 1}

The fugitives form a clear stream with the same orientation as the stream found
by \citet{2021A&A...645A..84M}, but we are able to find a few fugitives
farther from the cluster than anything they report.
The trailing arm of the stream seems to be much longer than the leading arm,
but it is not clear if this is because of the actual extent of the arms or is
this appearance the result of the very small number of identified fugitives. 

\subsubsection{IC 2391}

The fugitives seem to form a stream with an orientation roughly similar to that
found by \citet{2021A&A...645A..84M}.
However, the spatial extent of the fugitives is somewhat smaller than the
structure they found.

\subsubsection{NGC 2516}

The fugitives form a clear stream with an orientation very similar to the
stream found by \citet{2021A&A...645A..84M}.
The spatial extent of the bulk of the fugitives is somewhat smaller, but there
are a handful of stars with moderate fugitive probabilities closer to us and
farther from the cluster than the end of the stream \citet{2021A&A...645A..84M}
find.

\subsubsection{Pleiades}

The fugitives seem to form a clear stream, but differently from the other
clusters it is the inner arm of the stream that is trailing and the outer arm
that is leading.
This orientation matches the orientation \citet{2021A&A...645A..84M} find,
but their results do not include the inner trailing arm, whereas we find many
fugitives there.

\section{Discussion}\label{sect:discussion}

\begin{figure*}
    \centering
    \includegraphics[width=17cm]{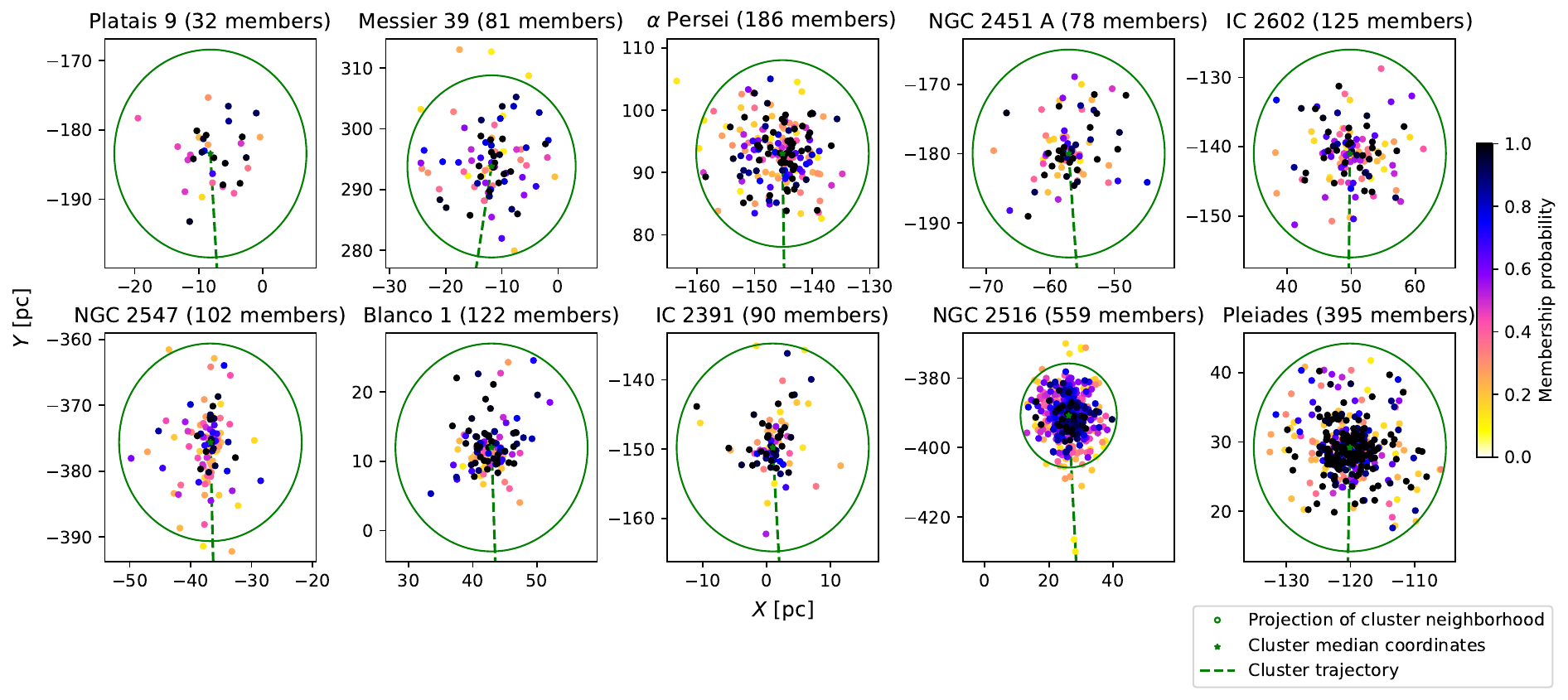}
    \includegraphics[width=17cm]{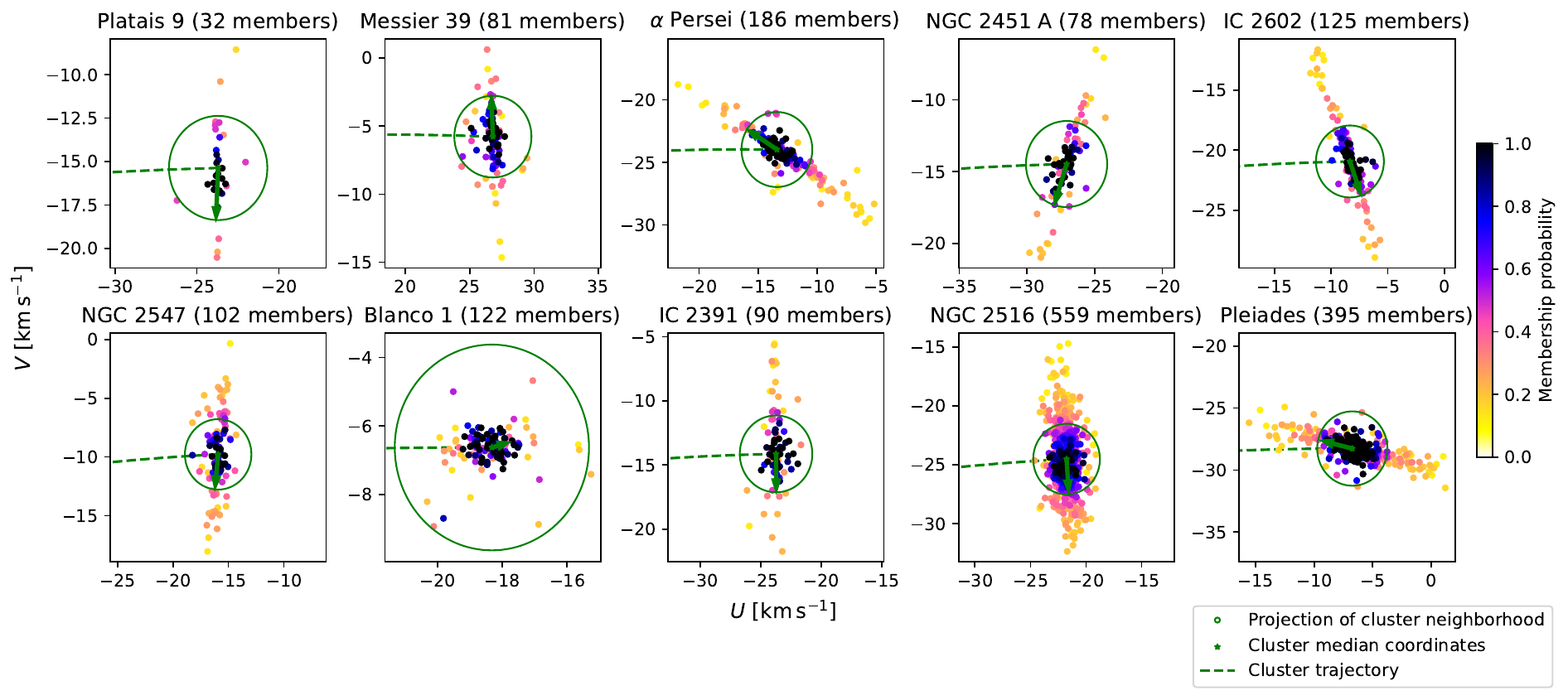}
    \caption{
	      Current-day position (top ten) and velocity (bottom ten) distribution
        of the members projected on the galactic plane.
        The green arrows show what a cluster's velocity components would be if 
        its radial velocity were larger by $3\,\mathrm{km\,s^{-1}}$ and
        therefore indicate the line-of-sight directions for each cluster.
        From the fact that the clusters appear to be strongly elongated along
        the line of sight in velocity space while not being visibly elongated in
        position space, it is apparent that the radial velocity uncertainty is
        much larger than the uncertainties of other phase space coordinates.
        \object{Blanco 1} is at a very high galactic latitude, so the apparent
        cluster elongation in velocity space is mostly along the $W$ axis and
        is not prominent in the $U-V$ plot.
    }
    \label{fig:members_current_coordinates}
\end{figure*}

\subsection{Radial velocity as the dominant source of uncertainty}
\label{subsect:rv_uncertainty_dominance}

Figure~\ref{fig:members_current_coordinates} shows the cluster member
coordinates in the $XY$- and $UV$-planes.
It can be seen that all clusters are strongly elongated along the line of
sight in velocity space.
This means that many stars have nominal radial velocities that place them
outside the cluster neighborhood, but the radial velocity uncertainty is large
enough that they are placed inside the neighborhood in a large fraction of the
Monte Carlo samples described in Sect.~\ref{subsect:general_idea}.
The fact that the clusters are elongated only along the line of sight but not
tangentially to it means that the tangential velocity uncertainty is much
smaller.
Furthermore, no clear elongation is visible in position space.
This means that the radial velocity is the dominant source of uncertainty and
applying quality cuts based on commonly used criteria such as RUWE would not
improve our results.

\subsection{Probability threshold value, $p_\mathrm{min}$}
\label{subsect:p_min_discussion}

\begin{figure*}
	\centering
    \includegraphics[width=17cm]{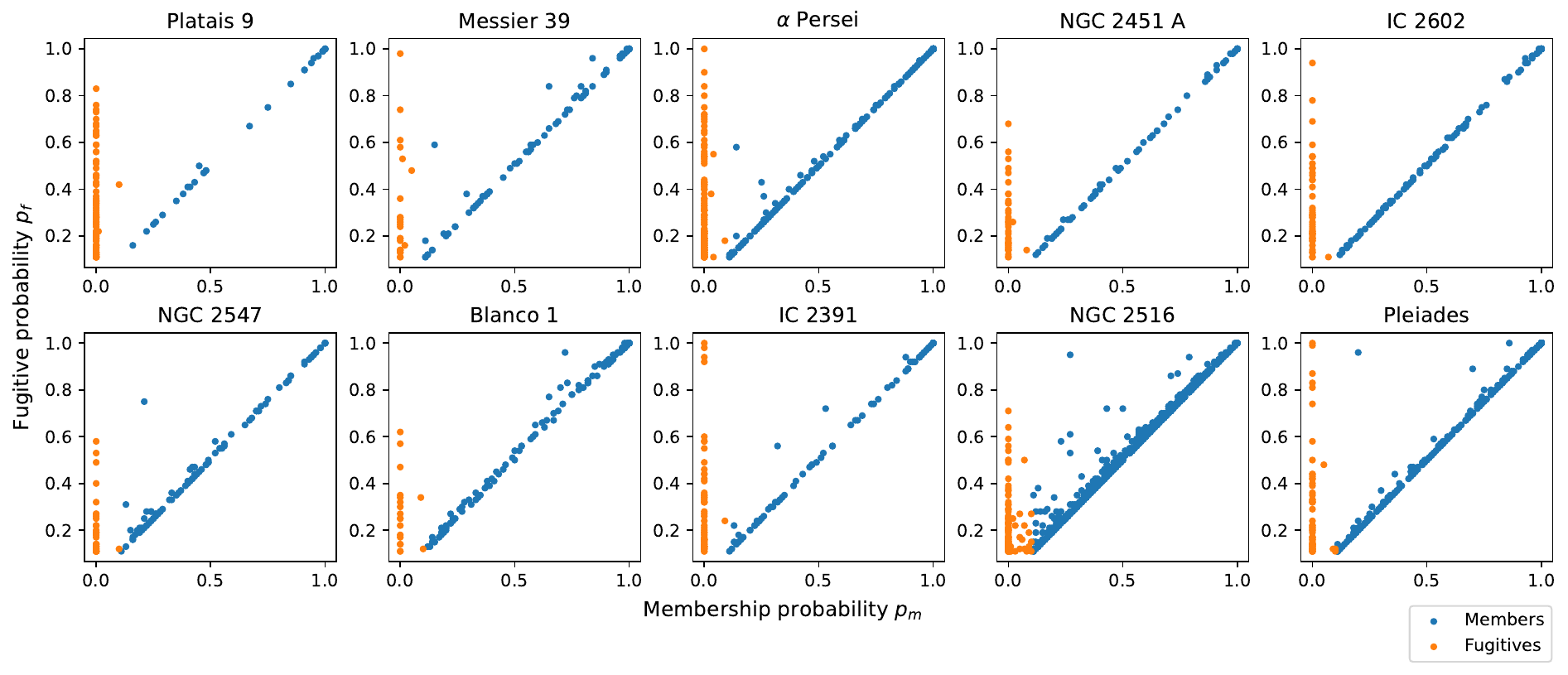} \caption{
        Comparison of fugitive and membership probabilities of members and
        fugitives.
        Stars close to the vertical line $p_m=0$ are clearly not members, and
        those close to the main diagonal are not fugitives.
        Stars in between could be classified as members or fugitives depending
        on the value of $p_\mathrm{min}$, but the number of such stars is very
        low.
        The only cluster where there are more than a handful of in-between stars
        is \object{NGC 2516}. This is due to the fact that it is the farthest
        away and, therefore, its stars have the most uncertain phase space
        coordinates. Even so, most of the ambiguous stars have low membership
        and fugitive probabilities.
    }
	\label{fig:limiting_probability_difference}
\end{figure*}

Our definitions of fugitives and members are meant to both exclude unrelated
field stars and also to distinguish stars currently very close to a cluster in
phase space, which can be associated with the cluster without traceback
computations, from the more distant stars whose connection to the cluster is
not obvious.
We used a single parameter for dividing the stars into these three groups with a
somewhat arbitrary value of $p_\mathrm{min}=0.1$, and using a different value
might reclassify some of our members as fugitives or some of our fugitives as
members.
That being said, there are two groups that are simple to interpret regardless of
the adopted value of $p_\mathrm{min}$.
First, stars with $p_m=0$ cannot be considered to be members.
Second, stars with $p_f=p_m$ cannot be classified as fugitives because either
$p_m > p_\mathrm{min}$ and they qualify as members, or $p_\mathrm{min} \geq p_f$
and they qualify as unrelated stars.
Stars with $p_f > p_m > 0$ could be classified as members or fugitives depending
on the adopted probability threshold value $p_\mathrm{min}$.

We can see in Fig.~\ref{fig:limiting_probability_difference}, however, that
the majority of stars are either on or very close to the vertical line $p_m=0$,
meaning they cannot be cluster members, or on or very close to the main
diagonal $p_f=p_m$, meaning they cannot be considered fugitives.
The number of stars that do not clearly belong to either of the two groups is
very low, with the exception of \object{NGC 2516}, which means that whether
stars get classified as members or fugitives of the other nine clusters is not
sensitive to the exact value of $p_\mathrm{min}$.
\object{NGC 2516} is the most distant of the ten clusters we considered,
so the stars associated with it have the largest uncertainties in their phase
space coordinates, but even in that cluster the number of stars that could be
classified as members or fugitives is not too large, and most of those have low
membership and fugitive probabilities regardless.

\subsection{Size of the cluster neighborhood}
\label{subsect:neighborhood_size}

The extent of a star cluster in phase space is nonzero, and measurement
uncertainties make the cluster appear even more spread out.
Performing our analysis using a neighborhood that is too small would therefore
cause us to miss cluster members and fugitives.
A very large neighborhood, however, leads to a very large field star
contamination rate.
Our choice must reflect these two competing incentives, meaning it should be
large enough to find as many true fugitives as possible while keeping the false
positive rate at an acceptable level.

From Table~\ref{tab:cluster_results} we can see that with the neighborhood
radii we used, $15\,\mathrm{pc}$ and $3\,\mathrm{km\,s^{-1}}$, the
median false positive rate of the ten clusters is $6\,\%$ and the maximum is
$14\,\%$.
We consider these values to be acceptable and see no reason to use smaller
neighborhoods.
If we set the neighborhood radii to $20\,\mathrm{pc}$ and
$3\,\mathrm{km\,s^{-1}}$, then the median false positive rate of the
clusters is $11\,\%$ and the maximum $33\,\%$.
If the neighborhood radii are $15\,\mathrm{pc}$ and
$4\,\mathrm{km\,s^{-1}}$, then the median false positive rate is $15.5\,\%$ and
maximum $29\,\%$.
We consider the false positive rates to be too large to justify using the larger
neighborhood radii.

\subsection{Parameters of the orbit of the Sun}

\begin{table*}[t]
	\begin{center}
        \caption{Alternative parameters of the solar orbit, also used by
		\citet{2021A&A...645A..84M}.}
		\begin{tabular}{l l l l}
			\hline\hline
			Parameter & Symbol & Value & Reference \\
			\hline
			Sun's galactocentric distance & $R_\sun$ & $8122\,\mathrm{pc}$ & (1) \\
			Sun's height from midplane & $z_\sun$ & $20.8\,\mathrm{pc}$ & (2)\\
			Sun's galactocentric velocity & $(v_{R,\sun},v_{\phi,\sun},v_{z,\sun})$ & $(12.9,245.6,7.78)\,\mathrm{km\,s^{-1}}$ & (3)\\
			\hline
		\end{tabular}
		\tablebib{
		    (1)~\citet{2018A&A...615L..15G};
		    (2)~\citet{2019MNRAS.482.1417B};
			(3)~\citet{2018RNAAS...2..210D}.
		}
		\label{tab:alternative_solar_parameters}
	\end{center}
\end{table*}

\begin{figure*}
	\centering
    \includegraphics[width=17cm]{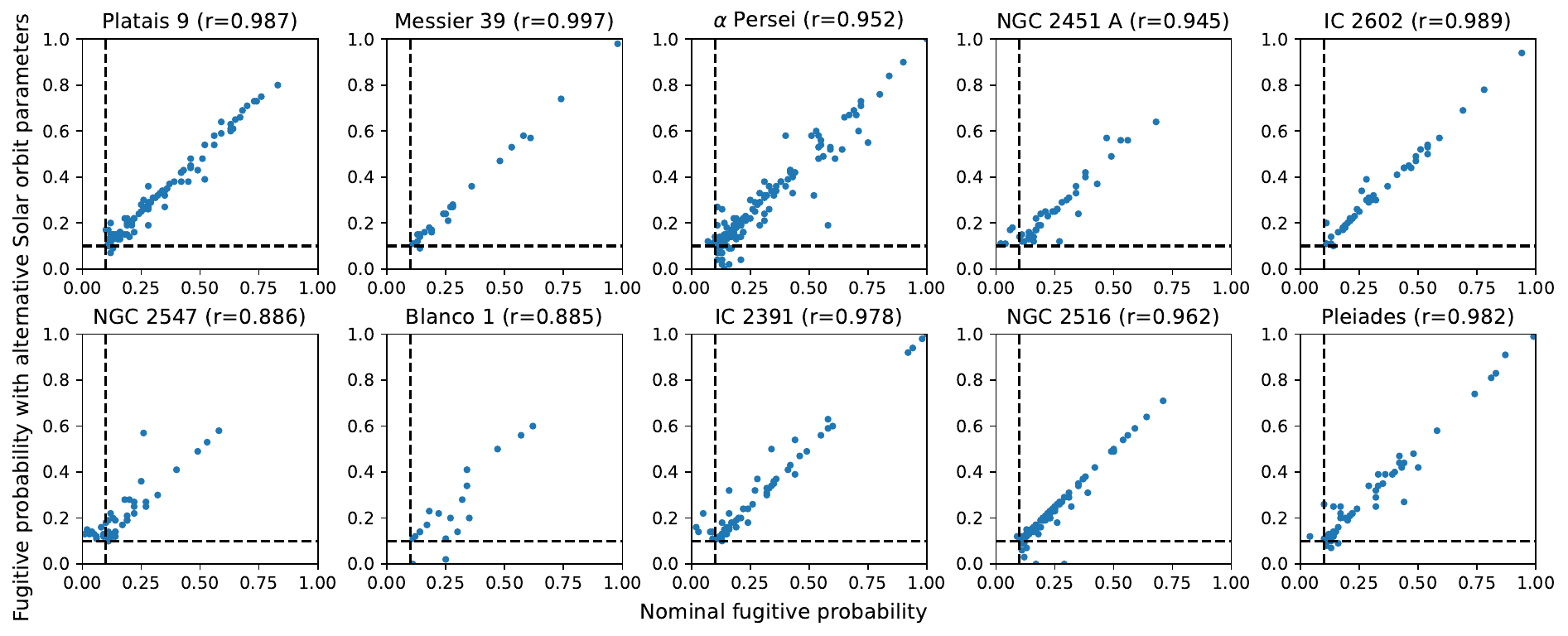}
    \caption{
        Comparison of fugitive probabilities computed with the nominal
        parameters of the orbit of the Sun, as listed in
        Table~\ref{tab:solar_parameters}, and with the alternative parameters 
        listed in Table~\ref{tab:alternative_solar_parameters}.
        Only stars with a fugitive probability $p_f > 0.1$ in either traceback
        and with a membership probability $p_m \leq 0.1$ are included.
        The Pearson correlation coefficients (in the panel titles) are close
        to 1, meaning the fugitive probabilities are not sensitive to the exact
        values of the solar orbit parameters.
    }
	\label{fig:different_solar_orbit}
\end{figure*}

Converting the heliocentric coordinates observed by \textit{Gaia} to
galactocentric coordinates used in the traceback computations requires knowing
the parameters of the orbit of the Sun.
Changing these parameters also changes the orbits of all other stars.
There are a few reasons to believe that our results should not be very sensitive
to such changes:
\begin{itemize}
    \item If a star and a cluster are on similar orbits with one set of solar
        orbit parameters, then they are also on similar orbits with some
        slightly different set of parameters.
    \item The time when a star enters the cluster neighborhood might change, but
        we are only interested in whether the star enters the neighborhood
        before some maximum traceback time or not.
    \item How close a star approaches a cluster might change, but we are only
        interested in whether the star enters the cluster neighborhood or not.
\end{itemize}
It is, however, important to verify that these assumptions indeed hold for our
data.
We therefore performed an additional traceback computation by assuming different
parameter values for the orbit of the Sun, listed in
Table~\ref{tab:alternative_solar_parameters}, with correspondingly different
galactocentric cylindrical phase space coordinates of the stars and clusters.
We did not modify the Galactic potential we used, but the change in the
galactocentric coordinates resulted in a change of the gravitational potential
at the clusters' locations.
We note that this is probably a worst-case scenario, as this choice means that
the Sun's velocity with respect to the local standard of rest changed by
${\sim}13\,\mathrm{km\,s^{-1}}$, with a similar effect for all the stars and
clusters we considered.
Figure~\ref{fig:different_solar_orbit} compares the fugitive probabilities
obtained from the two tracebacks.
The results are overall very similar, as demonstrated by the Pearson correlation
coefficient values (in the panel titles) being close to one.
We conclude that our results are indeed not sensitive to the values of the
parameters of the orbit of the Sun.

\subsection{Completeness and limitations}

In the \textit{Gaia} era it is the availability of radial velocities that limits
the completeness of the stellar sample that can be included in the traceback
computations.
Although it is possible to augment \textit{Gaia} radial velocities with other
catalogs, it would still remain true that the completeness would be lower for
fainter sources.
Furthermore, because completeness is a function of apparent magnitude, it is
more difficult to find fugitives that are behind their cluster (farther from us)
compared to the fugitives in front of it (closer to us).

Methods that do not require radial velocity measurements, such as the convergent
point method, suffer less from these drawbacks and therefore allow greater
completeness.
But some stars that the convergent point method would associate with a cluster
might be easily recognizable as unrelated field stars if radial velocity
measurements were available.
On the other hand, methods incorporating chemical tagging in addition to
kinematic data have the potential of producing fugitive samples with much lower
field star contamination rates than the traceback method, but they require
spectra with even higher quality than what is needed for radial velocity,
measurements.
The convergent point method, the traceback method and methods combining
kinematic data with chemical tagging therefore form a sequence with increasing
reliability at the cost of decreasing completeness.

It might also be advantageous to combine the methods.
Some version of the convergent point method could be used to relatively easily
exclude stars that are clearly kinematically distinct from the cluster.
Our traceback method could then be used to further refine the sample and to
quantify individual fugitive probabilities and the field star contamination
rate.
High-probability fugitives could in turn be good candidates for spectroscopic
followup that is required for strong chemical tagging.

\section{Summary}\label{sect:summary}

We present a new method of identifying stars that have escaped from nearby open
clusters in the last few ten million years, which we have applied to ten nearby
young clusters.
Our method is based on the assumptions that the fugitives have been close to the
cluster in phase space in the past, but we do not assume the fugitives are
close to the cluster at present.
We show how the fugitive probability of each individual star can be estimated
with a straightforward Monte Carlo procedure.
We also show how the field star contamination rate can be estimated for each
cluster by counting stars around a reference point, and describe how a suitable
reference point can be chosen.

Our method requires a model of the gravitational potential of the Galaxy and the
parameters of the orbit of the Sun.
They are constrained by observations, and we demonstrate that having exact
values is not critical.
Our method also requires choosing the size of the cluster neighborhood in phase
space.
Rough values are suggested by observations of open clusters, and the final
choice can be evaluated using the field star contamination rate.
We additionally adopted a probability threshold to avoid cluttering our analysis
with stars with very low fugitive probabilities, and although we also used the
same threshold value to distinguish cluster members from fugitives, we show that
this division does not strongly depend on the exact value of the threshold.

Our method is able to provide lower bounds of cluster ages.
The bounds we obtain with \textit{Gaia}~DR3 data are not very restrictive, but
all are consistent with the isochrone ages from \citet{2019A&A...623A.108B}.
Furthermore, the fugitives we identify are consistent with the cluster
isochrones on the HR diagram.
The spatial distribution of the fugitives we identify seems to be consistent
with the structures found by \citet{2021A&A...645A..84M}.
In cases where the extended stellar structures they found had a convincing
preferred orientation, we also tend to see an elongated structure with a very
similar orientation.

A shortcoming of our method is that it is reliant on \textit{Gaia} radial
velocity measurements.
The relatively small number of stars with such measurements limits the
completeness of our fugitive sample.
Additionally, for stars that do have such measurements available, the radial
velocity uncertainty tends to be the main contributor of their phase space
coordinate uncertainty, which limits the useful maximum traceback time.
Despite that, our method is already capable of complementing the fugitive lists
obtained with other methods and can be applied to many clusters relatively
easily.
Future \textit{Gaia} data releases are expected to include more stars with
radial velocity measurements and to have smaller uncertainties for the stars
with measurements available already.
Our method will be able to take advantage of such richer data sets with only
trivial modifications.

\begin{acknowledgements}
We thank the anonymous referee for helpful comments.
PM gratefully acknowledges support from project grants from the Swedish Research Council (Vetenskapr\aa det, Reg: 2017-03721; 2021-04153). 
Some of the computations in this project were completed on computing equipment bought with a grant from The Royal Physiographic Society in Lund.
This research has made use of NASA’s Astrophysics Data System.
This work made use of \texttt{astropy}: a community-developed core Python
package and an ecosystem of tools and resources for astronomy
\citep{astropy:2013, astropy:2018, astropy:2022}.
This work made use of \texttt{matplotlib} \citep{Hunter:2007} and \texttt{numpy}
\citep{2020NumPy-Array}.
\end{acknowledgements}

\bibliographystyle{aa}

\begin{appendix}

\section{Rough cluster coordinates}

Rough values of the cluster coordinates queried from SIMBAD
\citep{2000A&AS..143....9W} using \texttt{astroquery}
\citep{2019AJ....157...98G} are listed in
Table~\ref{tab:rough_cluster_coordinates}.
These values were used as initial guesses for the procedure described in
Sect.~\ref{subsect:clusterparameters} to obtain the nominal cluster
coordinates, so their exact values are not critical.

\begin{table}[h]
\centering
\caption{\label{tab:rough_cluster_coordinates}
Rough values of the cluster coordinates.}
\begin{tabular}{lcccccc}
\hline \hline
Cluster name & $\alpha$ & $\delta$ & $\varpi$ & $\mu_{\alpha*}$ & $\mu_\delta$ & $v_r$ \\
 &  &  & $\mathrm{mas}$ & $\mathrm{mas\,yr^{-1}}$ & $\mathrm{mas\,yr^{-1}}$ & $\mathrm{km\,s^{-1}}$ \\
\hline
\object{Platais 9} & $9^{\mathrm{h}}16^{\mathrm{m}}58^{\mathrm{s}}$ (1) & $-43^\circ51{}^\prime42{}^{\prime\prime}$ (1) & 5.5 (1) & -24.7 (1) & 13.3 (1) & 17.0 (2) \\
\object{Messier 39} & $21^{\mathrm{h}}31^{\mathrm{m}}33^{\mathrm{s}}$ (1) & $+48^\circ14{}^\prime48{}^{\prime\prime}$ (1) & 3.3 (3) & -7.4 (3) & -19.6 (3) & -5.4 (4) \\
\object{$\alpha$ Per cluster} & $3^{\mathrm{h}}26^{\mathrm{m}}28^{\mathrm{s}}$ (1) & $+48^\circ58{}^\prime30{}^{\prime\prime}$ (1) & 5.7 (3) & 22.9 (3) & -25.6 (3) & 0.2 (5) \\
\object{NGC 2451 A} & $7^{\mathrm{h}}42^{\mathrm{m}}57^{\mathrm{s}}$ (1) & $-38^\circ15{}^\prime48{}^{\prime\prime}$ (1) & 5.2 (1) & -21.1 (1) & 15.3 (1) & 27.7 (4) \\
\object{IC 2602} & $10^{\mathrm{h}}42^{\mathrm{m}}27^{\mathrm{s}}$ (1) & $-64^\circ25{}^\prime36{}^{\prime\prime}$ (1) & 6.6 (3) & -17.8 (3) & 10.7 (3) & 17.4 (3) \\
\object{NGC 2547} & $8^{\mathrm{h}}09^{\mathrm{m}}52^{\mathrm{s}}$ (6) & $-49^\circ10{}^\prime35{}^{\prime\prime}$ (6) & 2.5 (3) & -8.6 (3) & 4.3 (3) & 14.1 (2) \\
\object{Blanco 1} & $0^{\mathrm{h}}03^{\mathrm{m}}25^{\mathrm{s}}$ (1) & $-29^\circ57{}^\prime30{}^{\prime\prime}$ (1) & 4.2 (3) & 18.7 (3) & 2.6 (3) & 5.8 (3) \\
\object{IC 2391} & $8^{\mathrm{h}}41^{\mathrm{m}}10^{\mathrm{s}}$ (1) & $-52^\circ59{}^\prime30{}^{\prime\prime}$ (1) & 6.6 (3) & -24.9 (3) & 23.3 (3) & 14.6 (3) \\
\object{NGC 2516} & $7^{\mathrm{h}}58^{\mathrm{m}}06^{\mathrm{s}}$ (1) & $-60^\circ48{}^\prime00{}^{\prime\prime}$ (1) & 2.4 (3) & -4.7 (3) & 11.2 (3) & 24.2 (4) \\
\object{Pleiades} & $3^{\mathrm{h}}46^{\mathrm{m}}24^{\mathrm{s}}$ (1) & $+24^\circ06{}^\prime48{}^{\prime\prime}$ (1) & 7.4 (3) & 20.0 (3) & -45.5 (3) & 6.6 (5) \\
\hline
\end{tabular}
\tablebib{(1) \citet{2020A&A...633A..99C}; (2) \citet{2018A&A...619A.155S}; (3) \citet{2018A&A...616A..10G}; (4) \citet{2017A&A...600A.106C}; (5) \citet{2019A&A...623A..80C}; (6) \citet{2018MNRAS.481L..11B}.}
\end{table}

\section{Distribution of the fugitives on the sky}

\begin{figure}[h]
    \centering
    \includegraphics[width=17cm]{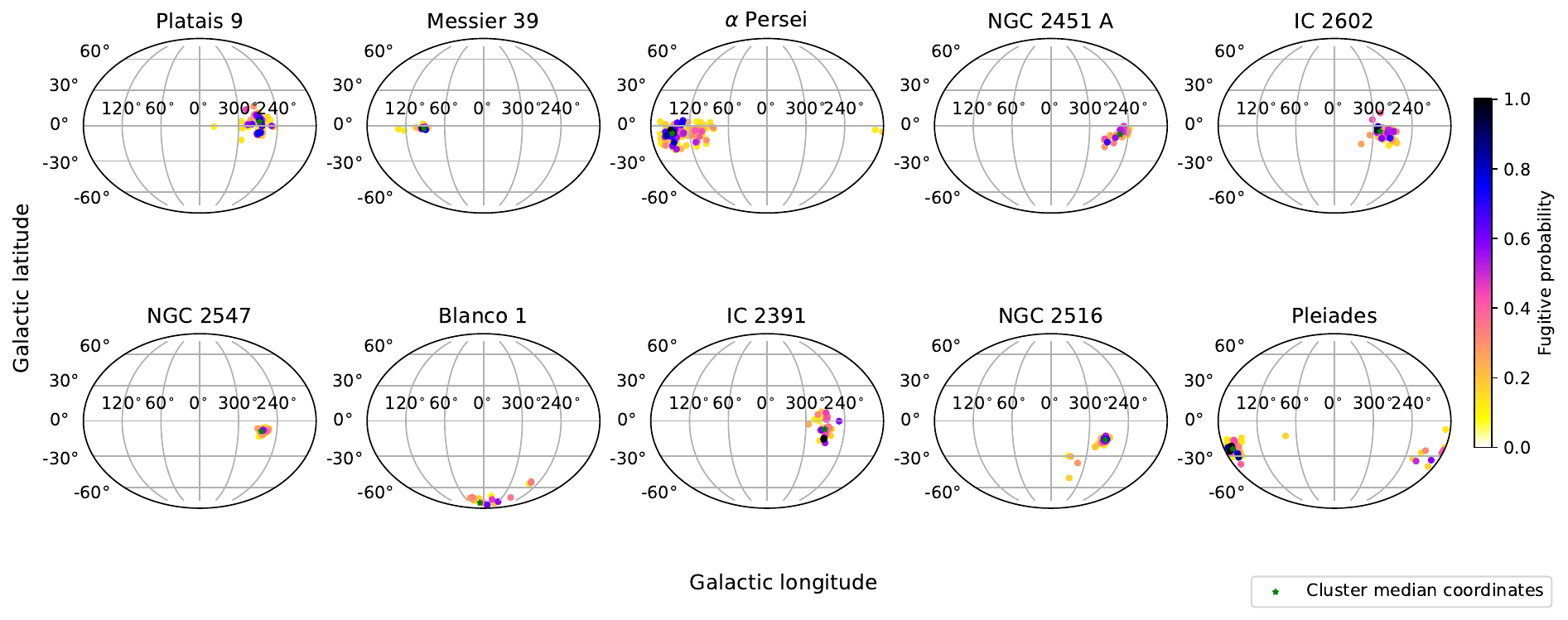}
    \caption{Current-day distribution of the fugitives on the sky.}
    \label{fig:SkyView}
\end{figure}

Figure~\ref{fig:SkyView} shows the distribution of the cluster fugitives on the
sky and demonstrates their large extent.

\end{appendix}

\end{document}